\documentclass[aps,prb,twocolumn,showpacs,superscriptaddress,nofootinbib,floatfix]{revtex4-1}

\usepackage{color}
\usepackage{amsmath}
\usepackage{amssymb}
\usepackage{graphicx}
\usepackage{bm}

\usepackage[version=4]{mhchem}
\usepackage{tikz}
\usetikzlibrary{calc}
\usetikzlibrary{graphs,trees,positioning,arrows,quotes}
\usepackage{pgfplots}
\pgfplotsset{compat=1.15}

\newcommand{\qsgw}{QS\emph{GW}}
\newcommand{\qsgwh}{QS\emph{G\^W}}

\begin{document}



\title{Multiple Slater determinants and strong spin-fluctuations as key ingredients of the electronic structure of electron- and hole-doped \ce{Pb_{10-x}Cu_{x}(PO4)6O}}

\author{Dimitar\,Pashov}
\affiliation{King's College London, Theory and Simulation of Condensed Matter, The Strand, WC2R 2LS London, UK}

\author{Swagata\,Acharya}
\affiliation{National Renewable Energy Laboratory, Golden, Colorado 80401}

\author{Stephan\,Lany}
\affiliation{National Renewable Energy Laboratory, Golden, Colorado 80401}

\author{Daniel S.\,Dessau}
\affiliation{Department of Physics, University of Colorado, Boulder, CO, 80309, USA}
\affiliation{Center for Experiments in Quantum Materials, University of Colorado Boulder, Boulder, CO 80309}
\affiliation{National Renewable Energy Laboratory, Golden, Colorado 80401}

\author{Mark\,van Schilfgaarde}
\affiliation{National Renewable Energy Laboratory, Golden, Colorado 80401}

\begin{abstract}

LK-99, with chemical formula \ce{Pb_{10-x}Cu_{x}(PO4)6O}, was recently reported to be a room-temperature
superconductor.  While this claim has met with little support in a flurry of ensuing work, a variety of calculations
(mostly based on density-functional theory) have demonstrated that the system possesses some unusual
characteristics in the electronic structure, in particular flat bands.  We have established previously that within DFT,
the system is insulating with many characteristics resembling the classic cuprates, provided the structure is not
constrained to the \emph{P}3(143) symmetry nominally assigned to it.  Here we describe the basic
electronic structure of LK-99 within self-consistent many-body perturbative approach, quasiparticle self-consistent \emph{GW} (\qsgw{}) approximation and their diagrammatic extensions.
\qsgw{} predicts that pristine LK-99 is indeed a Mott/charge transfer insulator, with a bandgap gap in excess of 3\,eV, whether or
not constrained to the \emph{P}3(143) symmetry.  The highest valence bands occur as a pair, and
look similar to DFT bands.  The lowest conduction band is an almost dispersionless state of largely Cu \emph{d}
character.  When \ce{Pb9Cu(PO4)6O} is hole-doped, the valence bands modify only slightly, and a hole pocket appears.  However, two
solutions emerge: a high-moment solution with the Cu local moment aligned parallel to neighbors, and a low-moment solution
with Cu aligned antiparallel to its environment.  In the electron-doped case the conduction band structure changes
significantly: states of mostly Pb character merge with the formerly dispersionless Cu \emph{d} state, and high-spin and
low spin solutions once again appear.  Thus we conclude that with suitable doping, the ground state of the system is not
adequately described by a band picture, and that strong correlations are likely.


\end{abstract}


\maketitle

\section{Introduction}
\label{sec:intro}

A recent paper~\cite{LK99} reported that Cu-doped \ce{Pb10(PO4)6O} (AKA LK-99, the trademark name
given by its discoverers) exhibits a Meissner effect up to 400\,K.  This rather startling claim~\cite{lee2023firs,lee2023consideration} has met with healthy
skepticism in the scientific community, with wide speculation that the levitation they reported was an artifact of some weak ferromagnetic interaction~\cite{guo2023ferromagnetic,timokhin2023p}.  If superconductivity at 400\,K were indeed confirmed, it would be a truly
remarkable and transformational discovery with major consequences for a wide range of technologies.  Unfortunately,
recent attempts to find superconductivity in LK-99 have been unsuccessful~\cite{liu2023semiconducting,kumar2023absence,kumar2023synthesis,guo2023ferromagnetic,timokhin2023p,wu2023successful},
damping the initial optimism. There have also been suggestions the presence of \ce{Cu2S} byproduct can explain the observed drop in resistivity~\cite{jain2023p}. Nevertheless this materials system is very interesting in its own right as a template
for hosting unusual one-particle properties, which as we show here are highly conducive to interesting and exotic
many-body effects.  In particular we establish, using a self-consistent form of \emph{ab initio} many-body
perturbation theory, the extremely flat valence and conduction bands are highly susceptible to strong correlations
owing to spin fluctuations.

Recent density-functional theory (DFT) calculations~\cite{griffin2023origin,lai2023first,RafalLK99,si2023electronic,cabezas-escares2023p,sun2023metallization,tao2023cu,jiang2023pb9cupo46oh2} agree that \ce{Pb10(PO4)6O} is a band insulator, however, there is little agreement on the metallic/insulating character of \ce{Pb9Cu(PO4)6O}. As we have established in Ref.~\cite{RafalLK99}, various flavors of DFT predict \ce{Pb9Cu(PO4)6O} to be a Mott/charge transfer insulator, provided the system is relaxed and not constrained to its nominal \emph{P}3 symmetry, a crucial ingredient of the electronic structure of LK-99 that most other DFT calculations missed.  This~\cite{RafalLK99}
work also showed the gap to be robust against the most likely kinds of disorder that preserve stoichiometry.  Here we
use a higher level of theory, the Quasiparticle Self-Consistent \emph{GW} (\qsgw{})~\cite{mark06qsgw,questaal_paper} approximation, and affirm the
principal result of that work: the pristine system is insulating and the gap is larger than found from
density-functionals \cite{cabezas-escares2023p}.
It also turns out that relaxation of the \emph{P}3 is not at all essential: the energy band structures of the
fully relaxed structures and structures constrained to the \emph{P}3 symmetry are very similar. Recent synthesis of phase-pure single crystals~\cite{puphal2023single} of \ce{Pb9Cu(PO4)6O} shows the system to be an insulator and optically transparent, a conclusion in complete consistency with our previous~\cite{RafalLK99} and present studies.   


Our study turns to the effect of electron- and hole-doping of \ce{Pb9Cu(PO4)6O}.  Indeed, some remarkable many-body effects emerge in both
cases.  In the hole doping case, the system can be reasonably approximated by a conventional picture: change in
occupation of largely unaltered valence bands.  However, two stable magnetic solutions emerge at the \qsgw{} level,
a high-spin state and a low-spin state.  Electron doping is more exotic because of the magnetic (Cu \emph{d}) character
of the lowest conduction band, as we explain below.  At the \qsgw{} level of theory the rigid band approximation
breaks down, which leads to somewhat exotic changes in the one-particle spectrum, with higher lying bands hybridizing
with the quasi-atomic Cu \emph{d} state, and give rise to a conduction band consisting several bands, derived mostly
from Cu and Pb.  As in the hole-doped case, a high-spin and a low-spin state can be stabilized.

The presence of two nearly degenerate spin states suggests that the ground state is a strongly correlated many-body
state that lies outside of a band picture, which is limited to a single Slater determinant.  In this picture, the
correlations arise from spin exchange between the Cu and O (hole doped case) and between Cu and Pb (electron doped
case).


\section{Lattice Structure}
\label{sec:structure}

The parent compound \ce{Pb9Cu(PO4)6O}, (AKA LK-99, the trademark name given by its discoverers), forms in the
hexagonal \emph{P}3 structure, space group 143.  Stoichiometrically, it may be regarded as combinations of formula units
\ce{(Pb3(PO4)2)3PbO}.
\begin{figure}
    \centering
    \includegraphics[scale=0.12]{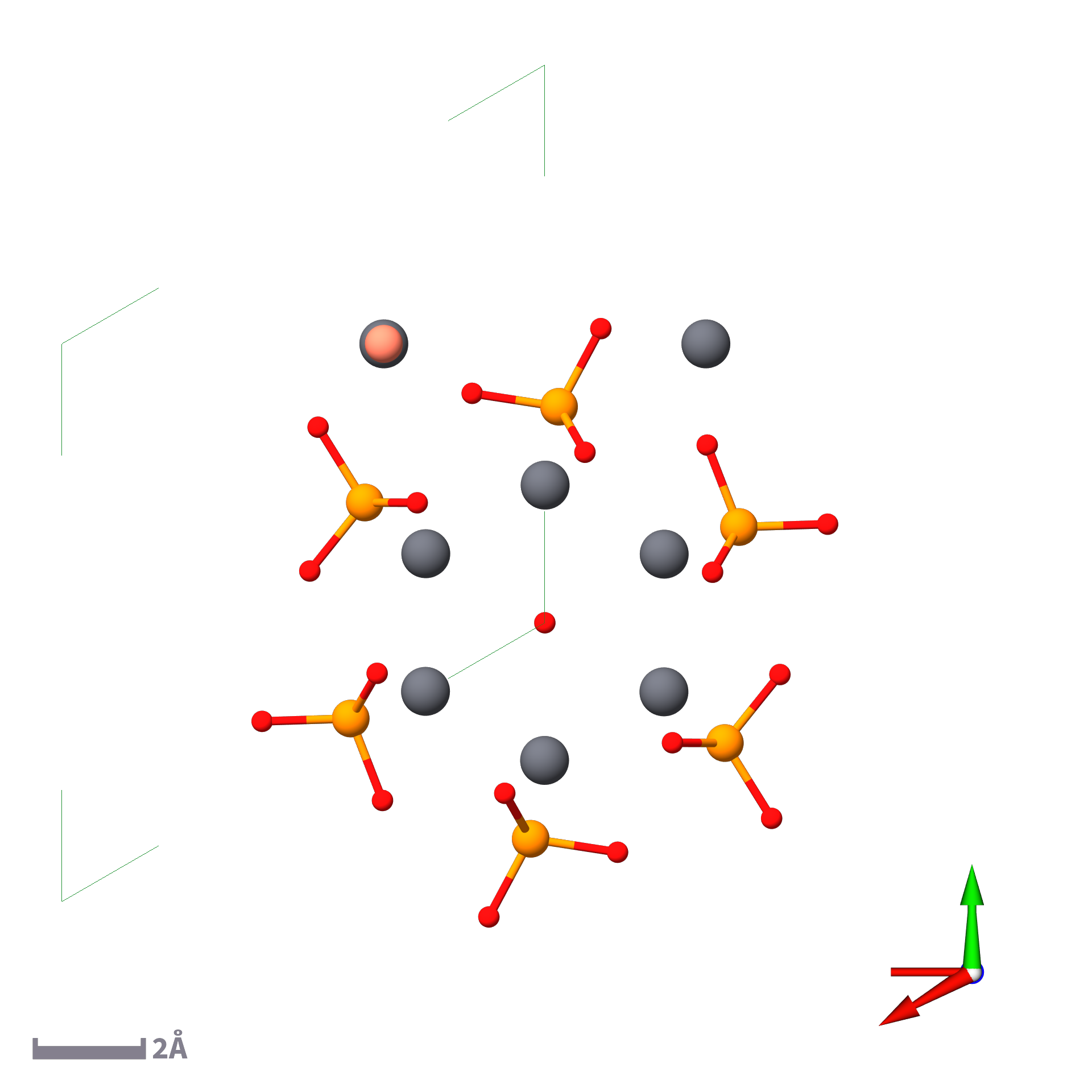} 
    \caption{Cartoon of the symmetrised LK-99 structure, viewed in direction of the \emph{z} axis. Large grey atoms are Pb, pink is Cu, yellow P and red is O.}
    \label{fig:structure}
\end{figure}

Six of the Pb (which may be called Pb(6\emph{h}) using Wykoff conventions) form a staggered set of triangles consistent
with three-fold rotational symmetry along the \emph{c} axis.  The other four Pb(4\emph{f}) atoms lie farther from this
axis.  Each P is surrounded by four almost equidistant O atoms, in a slightly deformed tetrahedron.  From the
perspective of chemical bonding, the P form $sp^{3}$ hybrids which bond to an O $p$ orbital and form bond-antibond
pairs.  Assigning a formal charge of +5 and $-$2 to P and O, the \ce{PO4} unit has a formal charge of $-3$.  Treating
the Pb \emph{s} states as core, Pb has two valence electrons, so \ce{(Pb3(PO4)2)} forms a combination of
covalent and ionic bonds.  PbO also forms an ionic bond; thus the parent compound \ce{(Pb3(PO4)2)3PbO} is
a mixed covalent-ionic insulator.  This last O sits in the center of the Pb rings and forms a linear chain on the
\emph{c}-axis.  Four sites are available to it, but only one site is occupied.  We use the minimum-energy configuration
as determined from the SCAN functional~\cite{sun_2015} (Ref.~\cite{RafalLK99}).  The Cu-O$_{|}$ distance is 5.78\,\AA.

\section{Energy band structure of the Parent LK-99 compound}
\label{sec:bands}

By the parent compound we refer to an ideal \ce{Pb9Cu(PO4)6O}, with O$_|$ occupying the lowest energy
configuration determined from the SCAN functional, as described in Ref.~\cite{RafalLK99}.  Relaxations consistent with
P3 were considered (symmetry-restricted), and also the system was allowed to fully relax (unrestricted). As noted in
that reference, the electronic structure does not depend on which site O$_|$ occupies (all four original O (4e) sites are equivalent. Therefore, the choice of which site to occupy does not matter in the 41 atom cell, however, non-equivalent O configurations exist in supercells).  Moreover, the Cu-Cu exchange interactions are predicted by SCAN to be much smaller than room temperature (as
might be expected in any case owing to the large Cu-Cu separation).  Thus, if the Cu have local moments as
density-functionals predict, the local moments will have no long range order.  Ref.~\cite{RafalLK99} also notes that the
electronic structure of both antiferrromagnetic and paramagnetic spin configurations are very similar to the
ferromagnetic one, provided in the FM case spin up and spin down are symmetrized.  We restrict consideration to the FM
case here, keeping in mind that to derive a good description of the electronic structure from it, the majority and
minority channels should be symmetrized.

For the single-particle band structure, we use the Quasiparticle Self-Consistent \emph{GW} approximation in the Questaal
package \cite{questaal_web,questaal_paper}.  Self-consistency is realized in a quasiparticle sense: the energy-dependent
self-energy includes many-body effects, especially plasmon excitations, but the quasiparticlization procedure replaces
the dynamical self-energy with a static one to yield an optimized single Slater determinant.  Quasiparticlization uses a
principle that minimizes some norm of the difference between the interacting $G$ and the noninteracting
one~\cite{mark06qsgw,Kotani07} and it also satisfies a variational principle~\cite{Beigi17}.  Importantly, it has the
property that the single-particle band structure should correspond to true excitation energies, in contrast to DFT.  Spin orbit coupling is included in these calculations.

\begin{figure}
    \begin{tikzpicture}
        \node [anchor=north west] (a) at (0,0)                    {\includegraphics[scale=0.35]{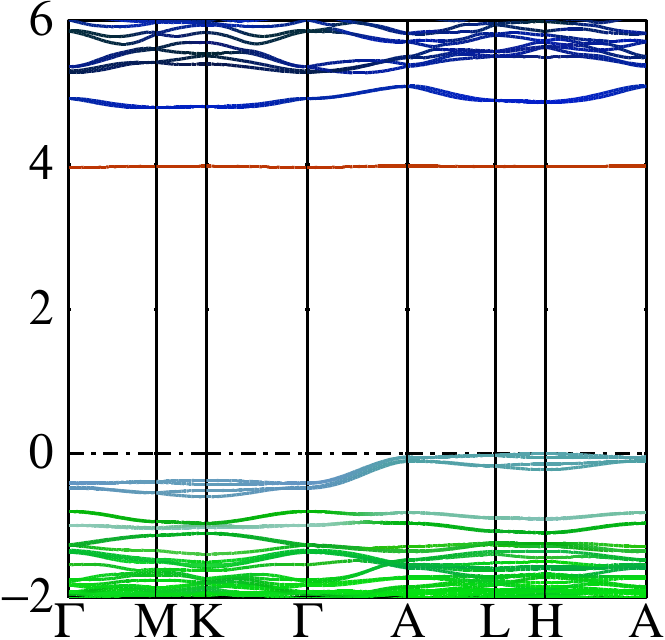}};
        \node [anchor=north west] (b) at ($(a.north east)+(0,0)$) {\includegraphics[scale=0.35]{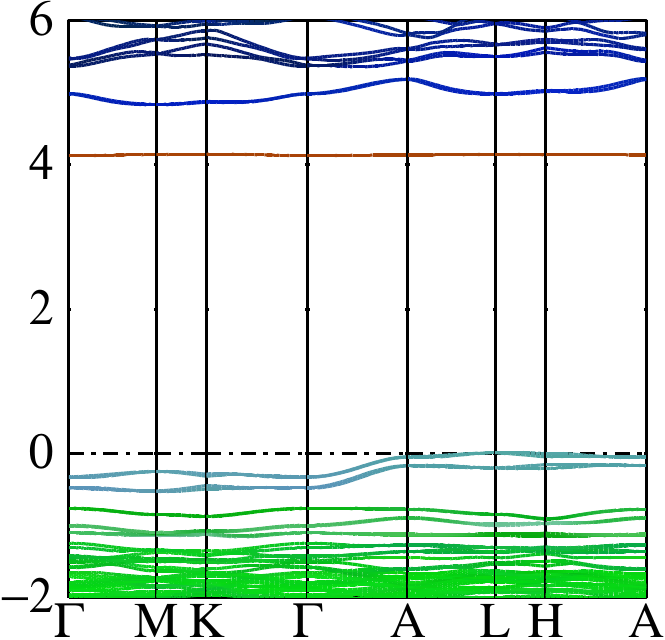}};
        \node [anchor=north west] (c) at ($(a.south west)+(0,0)$) {\includegraphics[scale=0.35]{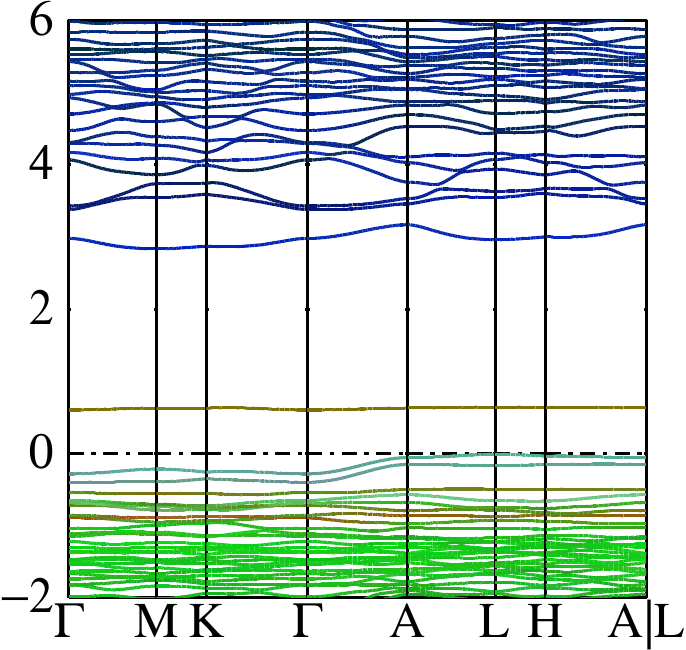}};
        \node [anchor=north west] (d) at ($(b.south west)+(0,0)$) {\includegraphics[scale=0.35]{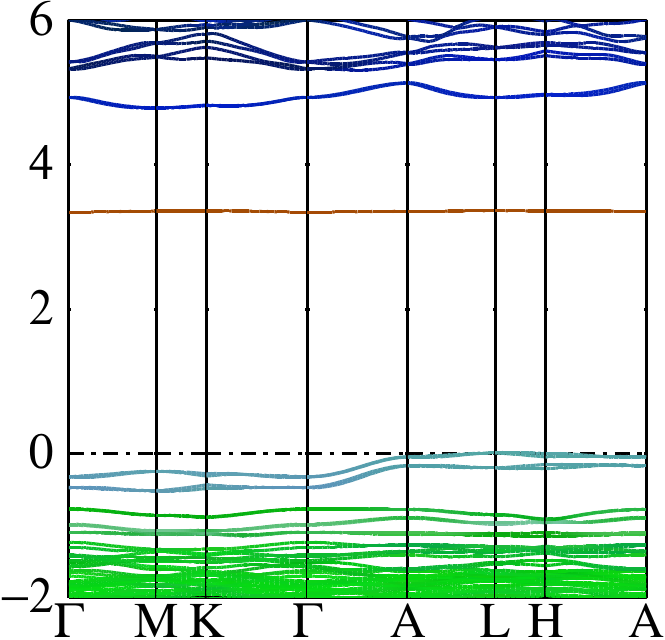}};
        \node [rotate=90] at ($(a.west)$) {$\mathrm{E - E_F,\ eV}$};
        \node [rotate=90] at ($(c.west)$) {$\mathrm{E - E_F,\ eV}$};
        \node at ($(a.east) + (-0.5,0.0)$) {(\emph{a})};
        \node at ($(b.east) + (-0.5,0.0)$) {(\emph{b})};
        \node at ($(c.east) + (-0.6,0.0)$) {(\emph{c})};
        \node at ($(d.east) + (-0.5,0.0)$) {(\emph{d})};
    \end{tikzpicture}
	\caption{Energy bands of the parent compound \ce{Pb9Cu(PO4)6O}.  Top panels show calculations in the QS$GW$ approximation for the
          symmetry-constrained relaxation (\emph{a}) and the fully relaxed case (\emph{b}).  Panel (\emph{c}) shows the
          band structure obtained from the PBESsol functional, using the fully relaxed structure and panel (\emph{d}) how
          the band structure is modified from \qsgw{} (\emph{a}) when ladders are added to the polarizability
          (\qsgwh{})~\cite{Cunningham2023}.  Red, green, and blue depict Mulliken projections onto Cu, O$_{P}$, and
          Pb orbitals.  Turquoise seen in the highest valence bands is a result of O$_{|}$ mixed in with O$_{P}$
          character.  All structural relaxations were obtained from the SCAN functional, and all energy bands include
          spin orbit coupling.}
	\label{fig:parent-bands}
\end{figure}
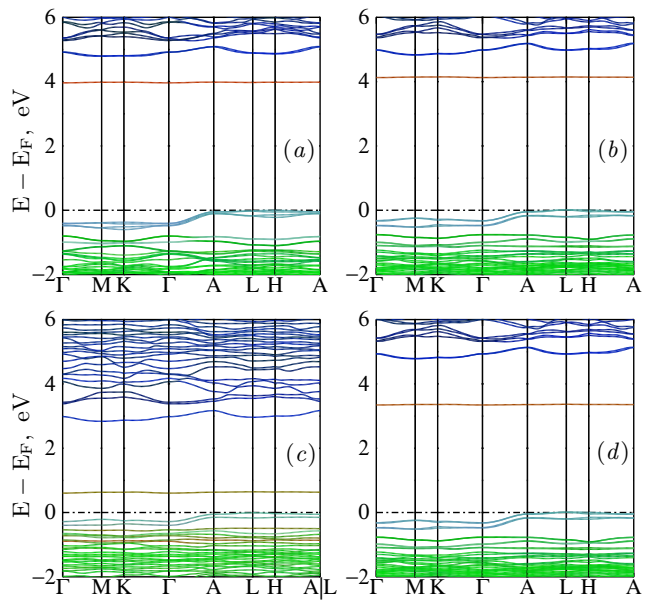

The energy bands of the parent compound (Fig.~\ref{fig:parent-bands}) neatly divide into a band O$_{P}$ \emph{p} states
well below the Fermi level $E_{F}$ (green), and pair of states of mixed O$_{P}$+O$_{|}$ character (turquoise) split off
from the other O-derived bands, and forming the valence band maximum.  Cu is $d^{9}$ with a lone Cu orbital forming a
nearly dispersionless conduction band minimum (red), and finally states of (mostly \emph{s}) Pb character sitting above
it (blue).  Top panels show bands for structures relaxed with the SCAN functional subject to a symmetry constraint
(\emph{a}) and fully relaxed.  There is very little difference in the two cases, except in constrained relaxation the
two bands at the valence band maximum are degenerate at A (apart from a slight lifting of degeneracies because of
majority and minority spins are not equivalent owing to the ferromagnetic approximation noted above).  In the
unrestricted relaxation, P3 symmetry is lifted and so is the degeneracy.

The \qsgw{} band structure has notable differences with those generated by standard density functionals.  In the
latter case, the symmetry-constrained structure is predicted to be metallic; only by breaking symmetry one Cu \emph{d}
state splits off to form the the conduction band.  Even after a gap forms under relaxation, the gap is small, and the
conduction band contains much more O character (compare the red color of the \qsgw{} conduction band to the
yellowish PBEsol~\cite{PBESOL} result).  Adding $U$ to PBE increases the Cu local moment, shifting and conduction band closer to the \qsgw{} result, the amount depending on the value of $U$.  However, even for $U{=}5$ the gap is still
${\sim}1.5$\,eV.  Also note that the gap between higher-lying (Pb-derived) bands and the valence band is much reduced.
This reflects a well-known tendency of DFT to underestimate bandgaps.  These differences will turn out to be important
when doping is considered.

On the other hand, \qsgw{} has a well-known tendency to overestimate bandgaps, largely because ladder diagrams are
omitted from the polarizability~\cite{Cunningham2023}.  This tendency is especially pronounced when levels are spin
split by exchange interactions between localized spins on a given site, e.g. in NiO, CoO, and \ce{La2CuO4}.  It has
been found that adding ladder diagrams largely ameliorates this tendency.~\cite{Cunningham2023} Extending \qsgw{} to
include ladder diagrams in \emph{W} we denote \qsgwh{}, and we consider it here to assess its effect on the ideal
LK99 parent compound.  The resulting energy band structure, Fig.~\ref{fig:parent-bands}(\emph{d}) looks quite similar to
\qsgw{}, except the unoccupied Cu state is shifted down by ${\sim}0.8$\,eV.  This reduction in the gap is similar
to, but slightly less than, what is found for NiO, CoO, CuO, \ce{Fe3O4}, and
\ce{La2CuO4}.~\cite{Cunningham2023} We also note that the position of this level is very sensitive to small changes
in the potential, and as a result the self-energy was difficult to stabilize.  But it hints at the extreme sensitivity
of the Cu-derived conduction band to small perturbations.

With Cu substituting for Pb, Cu \emph{d} states must lie at the Fermi level if restricted to nonmagnetic solutions.
Indeed calculations show that the Cu \emph{d} overlap with \emph{p} states belonging to O$_{|}$ and O$_{P}$.  Cu is
formally $d^{9}$, the valence band maximum has one hole, making the system metallic.  Bonds of both the Cu-\emph{d}
O-\emph{p} are have a dispersion of $\sim$0.2\,eV or less, and they weakly hybridize with each other.  Because of their
extremely low dispersion, their kinetic energy is small and both charge and spin easily fluctuate between Cu and the
environment.  It is noteworthy that any of common bosons responsible for many body effects are strong contenders in this
system. First, the electron-phonon interaction can be very strong because of the flat bands, and also because a
significant band of phonons involving O$_{P}$, with energies between 110-140\,meV, form part of the \ce{PO4}
network.  (Note that the O$_{|}$ phonons also have rather high energies, between 60-80\,meV). Second, charge fluctuations, e.g. between the weakly bonded but spatially
separated Cu and O$_{|}$ can be a source strong instabilities.  And finally, the spin fluctuations can be low-energy.
This will be particularly evident in the doped cases to be discussed next.

\section{Doping}

To avoid peculiarities of the ``chemical personality'' of specific dopants, we dope the system by adding a homogeneous
nuclear background charge, which must be compensated by adding or removing electrons to keep the unit cell neutral.  We
consider both electron and hole doping.  As check on the appropriateness of uniform background model, we compared the
electron-doped case to a system with a real dopant --- a F atom substituting for the chain-O, and find it similar to a
\qsgw{} calculation in the presence of a homogeneous background, requiring one extra electron.

\subsection{Hole-doped case}
\label{sec:hdope}

\begin{figure}
    \begin{tikzpicture}
        \node [anchor=north west] (a) at (0,0)              {\includegraphics[scale=0.32]{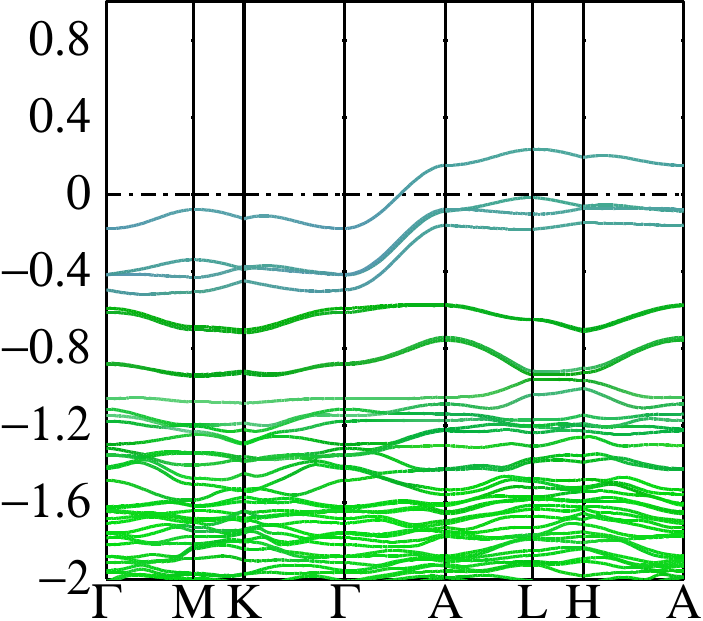}};
        \node [anchor=north west] (b) at ($(a.north east)$) {\includegraphics[scale=0.32]{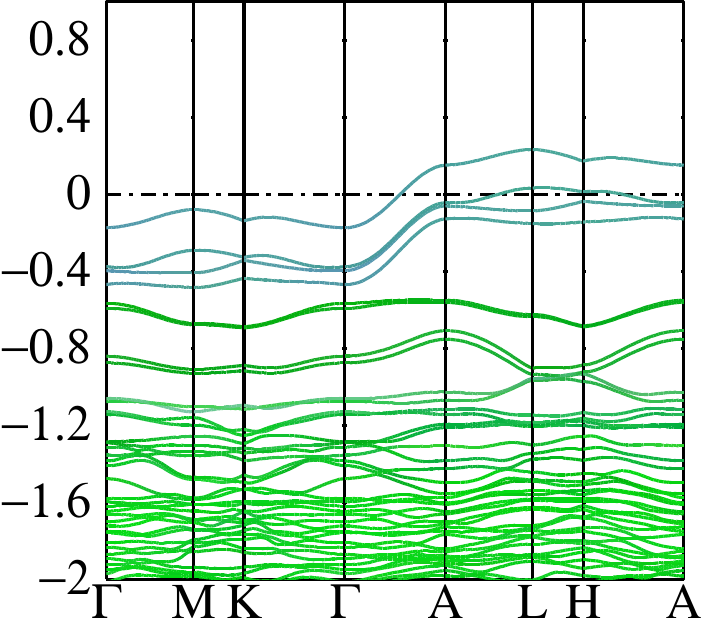}};
        \node [rotate=90] at ($(a.west)-(0.2,0)$) {$\mathrm{E - E_F,\ eV}$};
        \node at ($(a.north west) + (0.93,-0.4)$) {(\emph{a})};
        \node at ($(b.north west) + (0.93,-0.4)$) {(\emph{b})};
    \end{tikzpicture} 
	\caption{Hole doped \ce{Pb9Cu(PO4)6O}.  Left is low-spin solution, right high-spin solution}
	\label{fig:hole-doped}
\end{figure}

Doping LK-99 with holes acts similarly to doping cuprates or \emph{sp} semiconductors: the energy bands do not change
shape significantly and a hole pocket at the valence band maximum appears.  Fig.~\ref{fig:hole-doped} shows the
resulting energy band structure with doping of 1/2 electron per formula unit.  Perhaps the most surprising result to
emerge from \qsgw{} is that two self-consistent solutions can be stabilized with different spin configurations: a
``low-moment'' configuration with 0.5\,$\mu_B$ in the unit cell, and a ``high-moment'' configuration with 1.5\,$\mu_B$.
In both cases the Cu moment is very close to the undoped compound (0.81\,$\mu_B$), while the remaining moment is
distributed over O and Pb.  In the low-moment case, spins align antiferrromagnetically to the Cu on average, while they
align ferromagnetically in the high-moment case.  This strongly hints that the ground state cannot be comprised of a
single Slater determinant, and that spin fluctuations are inherently large.

\subsection{Electron-doped case}
\label{sec:edope}

\begin{figure}
    \begin{tikzpicture}
        \node [anchor=north west] (a) at (0,0)                      {\includegraphics[scale=0.33]{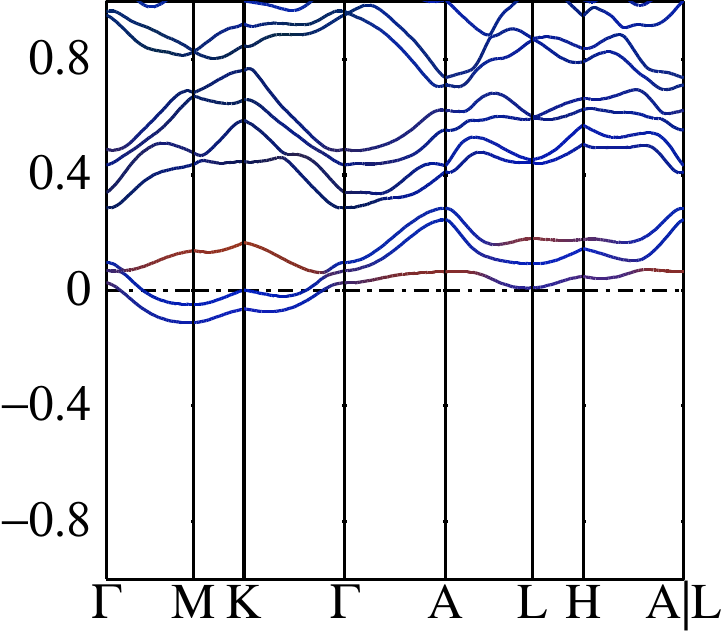}};
        \node [anchor=north west] (b) at ($(a.north east)-(0.2,0)$) {\includegraphics[scale=0.33]{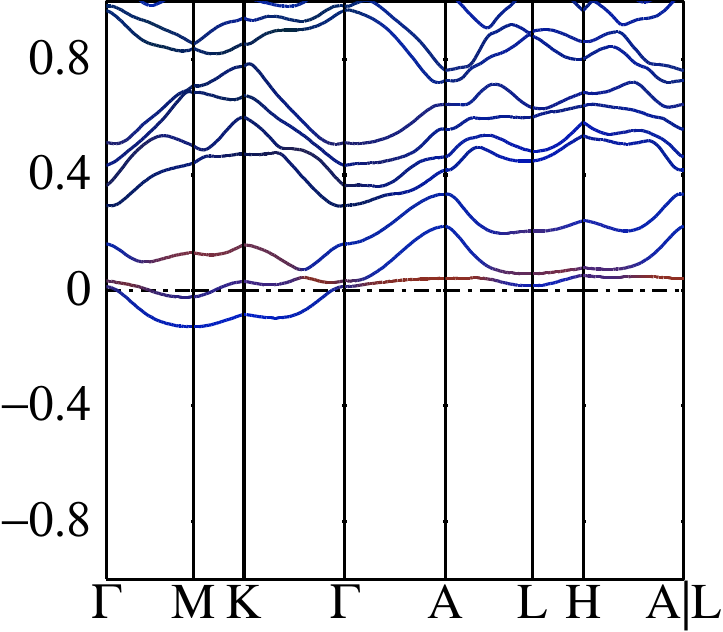}};
        \node [rotate=90] at ($(a.west)-(0.2,0)$) {$\mathrm{E - E_F,\ eV}$};
        \node at ($(a.south east) + (-0.6,0.65)$) {(\emph{a})};
        \node at ($(b.south east) + (-0.6,0.65)$) {(\emph{b})};
    \end{tikzpicture} 
	\caption{Electron doped \ce{Pb9Cu(PO4)6O}.  Left is low-spin solution, right high-spin solution}
	\label{fig:electron-doped}
\end{figure}

A very different picture emerges when LK-99 is doped with electrons.  The rigid-band approximation no longer applies.
For small doping, electrons occupy the unoccupied Cu \emph{d} state and reduce the local moment in proportion to the
doping.  When heavily doped, the splitting between states of Pb \emph{s} character and the lowest conduction band Cu
\emph{d} character can close.  This likely occurs because the gap between Cu and the valence band originates primarily
from an exchange splitting as a consequence of the onsite Hubbard interaction, while much of the splitting between the
unoccupied Pb(\emph{s}) and valence band is a consequence of the long-range screened coulomb interaction
$W$~\cite{Maksimov}.  The latter is important in determining the bandgap in \emph{sp} semiconductors, while the former
is largely responsible for the gap in transition metal oxides such as NiO, CoO, CuO and
\ce{La2CuO4}, where the gap closes if the local moment vanishes.  When the system is doped,
screening increases and the long-ranged part of $W$ is reduced.  However, the local screening on Cu (essentially on-site
$U$ and $J$) is much less affected.  When Pb-Cu splitting does close, the conduction band involves multiple states,
mixing the localized the Cu \emph{d} state and more dispersive Pb \emph{s} states (Fig.~\ref{fig:electron-doped}) Here
also a `low-moment'' configuration with 0.66\,$\mu_B$, and a ``high-moment'' configuration with 1.4\,$\mu_B$ can be
stabilized, with the Cu moment approximately similar to the parent compound. We note in passing DFT does not capture this effect, as it requires feedback between the screening and the band structure~\cite{Vidal10}.However, electron doping should be taken with some caution.  The wide gap mitigates against thermodynamically stable dopants, and it may not be  feasible to electron dope this system.

\section{Analysis and Conclusions}

The original work~\cite{RafalLK99} that established pristine LK-99 is insulating, in a DFT framework, has been confirmed
theoretically by several groups, using some form of DFT or a combination of DFT and DMFT~\cite{Korotin23,Si23}. This has led these groups to conclude that some form of doping that changes the electron count, is essential for LK-99 to be a metal.

The present work confirms that conclusion at a higher level of theory, and turns to the question of electron and hole
doping, within the \qsgw{} framework.  Two surprising results emerge from \qsgw{} that are not found in
density-functional theory. The first is the marked change in splitting between Pb states in the conduction band and the O-derived valence band states with doping.  This can be understood as a change in the screening as the system becomes metallic. The close relation between bandgaps in \textit{sp} bonded semiconductors and the long-range form of the screened coulomb interactions has been established for some time~\cite{Maksimov,Gruning06} and the effect of changes in screening on the band structure in such systems systems has been also been reported~\cite{Neaton06,Vidal10}. What is unique in the present case is that bandgaps with two different origins coexist: one of another of semiconductor character (Pb-O) which is sensitive to doping and another of Mott character (Cu-O) which is not.  The second remarkable finding is the coexistence of two spin states, with the Cu aligned either parallel or antiparallel to the host.  The presence of an atomic-like \textit{d} state in a semiconducting conduction band is also seen in Cu-doped ZnO.  But in the present case flat bands with low kinetic energy allow a moment to be induced on the \textit{sp} subspace that would not occur in Cu-doped ZnO, and the coexistence of two subspaces with different spin channels is new.    In the QS\textit{GW} approach, the quasiparticlization procedure demands a single Slater determinant; thus the alignment of the two kinds of spins
emerges as two distinct solutions.  While it was possible to stabilize two distinct spin configurations,
there was a strong tendency to flip between them as self-consistency proceeded.  This indicates that the spin
configurations are nearly degenerate in energy.  Self-consistency is clearly essential here: such a result would be
problematic with conventional forms of \emph{GW}, e.g. one-shot \emph{GW} based on DFT. The crucial role of 
 self-consistency has been discussed recently for several other insulating systems.~\cite{acharya2021importance,grzeszczyk2023strongly}

The near-degeneracy of distinct ground states in a one-particle description suggests that the true ground state cannot
be described by a single-determinant; and moreover correlations from spin fluctuations between the Cu local moment and
delocalized environment are likely to be very strong.  This does not by itself indicate a strong propensity for
superconductivity, but it does suggest a novel path to strong correlations, which may result in superconductivity. We believe, irrespective of whether room-temperature superconductivity is realized in these classes of materials or not, the fact that \ce{Pb10(PO4)6O} is a band insulator, \ce{Pb9Cu(PO4)6O} is a Mott insulator and the electron- and hole-doped \ce{Pb9Cu(PO4)6O} can not be described by a single Slater determinant, makes these class of systems a unique playground to explore the transition from band- to Mott-insulator to multi-determinantal nature of weakly doped Mott physics. 


\section{Methods}
We apply quasiparticle self-consistent $GW$ theory (\qsgw{}) \cite{mark06qsgw}, which, in contrast to conventional $GW$ methods, modifies the charge density and is determined by a variational principle \cite{Beigi17}. Further, \qsgwh{} \cite{Cunningham2023} is a diagrammatic extension of \qsgw{} where the screened coulomb interaction $W$ is computed including excitonic vertex corrections (ladder diagrams) by solving a Bethe–Salpeter equation (BSE) within Tamm-Dancoff approximation \cite{hirata1999}. Crucially, both our \qsgw{} and \qsgwh{} methods are fully self-consistent in both self-energy $\Sigma$ and the charge density \cite{acharya2021importance}. $G$, $\Sigma$, and $\hat{W}$ are updated iteratively until all of them converge. Such approaches are parameter-free and have no starting point bias.

\section{Acknowledgments}

This work was supported by National Renewable Energy Laboratory, operated by Alliance for Sustainable Energy, LLC,
for the U.S. Department of Energy (DOE) under Contract No. DE-AC36-08GO28308, funding from Office of Science, Basic
Energy Sciences, Division of Materials.  We also acknowledge the use of the Eagle facility at NREL, sponsored by the
Office of Energy Efficiency and also the National Energy Research Scientific Computing Center, under Contract
No. DE-AC02-05CH11231 using NERSC award BES-ERCAP0021783.  DD was supported by the
  U.S. Department of Energy, Office of Science, Office of Basic Energy Sciences, under Grant No. DE-FG02-03ER46066, and
  the Gordon and Betty Moore Foundation’s EPiQS Initiative through Grant No. GBMF9458.


\begin{thebibliography}{10}
	
	\bibitem{LK99}
	S.~Lee, J.~Kim, H.-T. Kim, S.~Im, S.~An, and K.~H. Auh, ``{Superconductor
		Pb$_{10-x}$ Cu$_{x}$(PO$_{4}$)$_{6}$O showing levitation at room temperature
		and atmospheric pressure and mechanism}.'' Preprint arXiv 2307.12037.pdf,
	2023.
	
	\bibitem{lee2023firs}
	S.~Lee, J.-H. Kim, and Y.-W. Kwon, ``The firs room-temperature ambient-pressure
	superconductor,'' {\em arXiv preprint arXiv:2307.12008}, 2023.
	
	\bibitem{lee2023consideration}
	S.~Lee, J.~Kim, S.~Im, S.~An, Y.-W. Kwon, and A.~K. Ho, ``Consideration for the
	development of room-temperature ambient-pressure superconductor (lk-99),''
	{\em Journal of the Korean Crystal Growth and Crystal Technology}, vol.~33,
	no.~2, pp.~61--70, 2023.
	
	\bibitem{guo2023ferromagnetic}
	K.~Guo, Y.~Li, and S.~Jia, ``Ferromagnetic half levitation of lk-99-like
	synthetic samples,'' {\em arXiv preprint arXiv:2308.03110}, 2023.
	
	\bibitem{timokhin2023p}
	I.~Timokhin, C.~Chen, Z.~Wang, Q.~Yang, and A.~Mishchenko, ``{Synthesis and
		characterisation of LK-99}.'' Preprint arXiv 2308.03823v2, 2023.
	
	\bibitem{liu2023semiconducting}
	L.~Liu, Z.~Meng, X.~Wang, H.~Chen, Z.~Duan, X.~Zhou, H.~Yan, P.~Qin, and
	Z.~Liu, ``Semiconducting transport in pb10-xcux (po4) 6o sintered from pb2so5
	and cu3p,'' {\em arXiv preprint arXiv:2307.16802}, 2023.
	
	\bibitem{kumar2023absence}
	K.~Kumar, N.~K. Karn, Y.~Kumar, and V.~P.~S. Awana, ``Absence of
	superconductivity in lk-99 at ambient conditions,'' 2023.
	
	\bibitem{kumar2023synthesis}
	K.~Kumar, N.~K. Karn, and V.~P.~S. Awana, ``Synthesis of possible room
	temperature superconductor lk-99: Pb9cu (po4) 6o,'' {\em Superconductor
		Science and Technology}, 2023.
	
	\bibitem{wu2023successful}
	H.~Wu, L.~Yang, B.~Xiao, and H.~Chang, ``Successful growth and room temperature
	ambient-pressure magnetic levitation of lk-99,'' {\em arXiv preprint
		arXiv:2308.01516}, 2023.
	
	\bibitem{jain2023p}
	P.~K. Jain, ``{Phase transition of copper (I) sulfide and its implication for
		purported superconductivity of LK-99}.'' Preprint arXiv 2308.05222v1, 2023.
	
	\bibitem{griffin2023origin}
	S.~M. Griffin, ``Origin of correlated isolated flat bands in copper-substituted
	lead phosphate apatite,'' 2023.
	
	\bibitem{lai2023first}
	J.~Lai, J.~Li, P.~Liu, Y.~Sun, and X.-Q. Chen, ``First-principles study on the
	electronic structure of pb10- xcux (po4) 6o (x= 0, 1),'' {\em Journal of
		Materials Science \& Technology}, 2023.
	
	\bibitem{RafalLK99}
	R.~Kurleto, S.~Lany, D.~Pashov, S.~Acharya, M.~van Schilfgaarde, and D.~S.
	Dessau, ``{Pb-apatite framework as a generator of novel flat-band CuO based
		physics, including possible room temperature superconductivity}.'' Preprint
	arXiv 2308.00698, 2023.
	
	\bibitem{si2023electronic}
	L.~Si and K.~Held, ``Electronic structure of the putative room-temperature
	superconductor pb $ \_9 $ cu (po $ \_4 $) $ \_6 $ o,'' {\em arXiv preprint
		arXiv:2308.00676}, 2023.
	
	\bibitem{cabezas-escares2023p}
	J.~Cabezas-Escares, N.~Barrera, C.~Cardenas, and F.~Munoz, ``Theoretical
	insight on the lk-99 material,'' {\em arXiv preprint arXiv:2308.01135}, 2023.
	
	\bibitem{sun2023metallization}
	Y.~Sun, K.-M. Ho, and V.~Antropov, ``Metallization and spin fluctuations in
	cu-doped lead apatite,'' 2023.
	
	\bibitem{tao2023cu}
	K.~Tao, R.~Chen, L.~Yang, J.~Gao, D.~Xue, and C.~Jia, ``The cu induced
	ultraflat band in the pb$_{10-x}$cu$_x$(po$_4$)$_6$o$_4$ ($x=0,0.5$),'' 2023.
	
	\bibitem{jiang2023pb9cupo46oh2}
	Y.~Jiang, S.~B. Lee, J.~Herzog-Arbeitman, J.~Yu, X.~Feng, H.~Hu, D.~Călugăru,
	P.~S. Brodale, E.~L. Gormley, M.~G. Vergniory, C.~Felser, S.~Blanco-Canosa,
	C.~H. Hendon, L.~M. Schoop, and B.~A. Bernevig, ``Pb$_9$cu(po4)$_6$(oh)$_2$:
	Phonon bands, localized flat band magnetism, models, and chemical analysis,''
	2023.
	
	\bibitem{mark06qsgw}
	M.~van Schilfgaarde, T.~Kotani, and S.~Faleev, ``{Quasiparticle Self-Consistent
		\emph{GW} Theory},'' {\em Phys. Rev. Lett.}, vol.~96, no.~22, p.~226402,
	2006.
	
	\bibitem{questaal_paper}
	D.~Pashov, S.~Acharya, W.~R.~L. Lambrecht, J.~Jackson, K.~D. Belashchenko,
	A.~Chantis, F.~Jamet, and M.~van Schilfgaarde, ``{Questaal: a package of
		electronic structure methods based on the linear muffin-tin orbital
		technique},'' {\em Comp. Phys. Comm.}, vol.~249, p.~107065, 2020.
	
	\bibitem{puphal2023single}
	P.~Puphal, M.~Y.~P. Akbar, M.~Hepting, E.~Goering, M.~Isobe, A.~A. Nugroho, and
	B.~Keimer, ``Single crystal synthesis, structure, and magnetism of
	pb$_{10-x}$cu$_x$(po$_4$)$_6$o,'' 2023.
	
	\bibitem{sun_2015}
	J.~Sun, A.~Ruzsinszky, and J.~P. Perdew, ``Strongly constrained and
	appropriately normed semilocal density functional,'' {\em Phys. Rev. Lett.},
	vol.~115, p.~036402, Jul 2015.
	
	\bibitem{questaal_web}
	\url{https://www.questaal.org}.
	\newblock {Questaal code website}.
	
	\bibitem{Kotani07}
	T.~Kotani, M.~van Schilfgaarde, and S.~V. Faleev, ``{Quasiparticle
		self-consistent \emph{GW} method: A basis for the independent-particle
		approximation},'' {\em PRB}, vol.~76, p.~165106, 2007.
	
	\bibitem{Beigi17}
	S.~Ismail-Beigi, ``{Justifying quasiparticle self-consistent schemes via
		gradient optimization in Baym–Kadanoff theory},'' {\em J. Phys.: Condens.
		Matter}, vol.~29, p.~385501, 2017.
	
	\bibitem{PBESOL}
	J.~P. Perdew, A.~Ruzsinszky, G.~I. Csonka, O.~A. Vydrov, G.~E. Scuseria, L.~A.
	Constantin, X.~Zhou, , and K.~Burke, ``Restoring the density-gradient
	expansion for exchange in solids and surfaces,'' {\em Phys. Rev. Lett.},
	vol.~100, p.~136406, 2008.
	
	\bibitem{Cunningham2023}
	B.~Cunningham, M.~Gr{\"u}ning, D.~Pashov, and M.~van Schilfgaarde,
	``{QS$G\hat{W}$: Quasiparticle Self Consistent $GW$ with Ladder Diagrams in
		$W$},'' {\em https://arxiv.org/abs/2302.06325}.
	
	\bibitem{Maksimov}
	E.~G. Maksimov, I.~I. Mazin, S.~Y. Savrasov, and Y.~A. Uspenski, ``Excitation
	spectra of semiconductors and insulators: a density-functional approach to
	many-body theory,'' {\em J. Phys.:Condens. Matter}, vol.~1, 1993.
	
	\bibitem{Vidal10}
	J.~Vidal, S.~Botti, P.~Olsson, J.-F. Guillemoles, and L.~Reining, ``{Strong
		Interplay between Structure and Electronic Properties in CuInSe$_{2}$: A
		First-Principles Study},'' {\em Phys. Rev. Lett.}, vol.~104, p.~056401, 2010.
	
	\bibitem{Korotin23}
	D.~M. Korotin, D.~Y. Novoselov, A.~O. Shorikov, V.~I. Anisimov, and A.~R.
	Oganov, ``{Electronic correlations in promising room-temperature
		superconductor Pb\textsubscript{9}Cu(PO\textsubscript{4})\textsubscript{6}O:
		a DFT+DMFT study}.'' Preprint arXiv 2308.04301, 2023.
	
	\bibitem{Si23}
	L.~Si, M.~Wallerberger, A.~Smolyanyuk, S.~di~Cataldo, J.~M. Tomczak, and
	K.~Held, ``{Pb$_{10-x}$Cu$_{x}$(PO$_{4}$)$_{6}$O: a Mott or charge transfer
		insulator in need of further doping for superconductivity}.'' Preprint arXiv
	2308.04427, 2023.
	
	\bibitem{Gruning06}
	M.~Gr{\"u}ning, A.~Marini, and A.~Rubio, ``{Density functionals from many-body
		perturbation theory: The band gap for semiconductors and insulators},'' {\em
		J. Chem. Phys.}, vol.~124, p.~154108, 2006.
	
	\bibitem{Neaton06}
	J.~B. Neaton, M.~S. Hybertsen, and S.~G. Louie, ``{Renormalization of Molecular
		Electronic Levels at Metal-Molecule Interfaces},'' {\em Phys. Rev. Lett.},
	vol.~97, p.~216405, 2006.
	
	\bibitem{acharya2021importance}
	S.~Acharya, D.~Pashov, A.~N. Rudenko, M.~R{\"o}sner, M.~van Schilfgaarde, and
	M.~I. Katsnelson, ``Importance of charge self-consistency in first-principles
	description of strongly correlated systems,'' {\em npj Computational
		Materials}, vol.~7, no.~1, pp.~1--8, 2021.
	
	\bibitem{grzeszczyk2023strongly}
	M.~Grzeszczyk, S.~Acharya, D.~Pashov, Z.~Chen, K.~Vaklinova, M.~van
	Schilfgaarde, K.~Watanabe, T.~Taniguchi, K.~Novoselov, M.~Katsnelson, {\em
		et~al.}, ``Strongly correlated exciton-magnetization system for optical spin
	pumping in crbr3 and cri3.,'' {\em Advanced Materials}, p.~2209513, 2023.
	
	\bibitem{hirata1999}
	S.~Hirata and M.~Head-Gordon, ``Time-dependent density functional theory within
	the tamm--dancoff approximation,'' {\em Chemical Physics Letters}, vol.~314,
	no.~3-4, pp.~291--299, 1999.
	
\end{thebibliography}


\end{document}